\documentclass{emulateapj}

\usepackage{subfigure}
\usepackage{url}
\usepackage[breaklinks,colorlinks,
  urlcolor=blue,citecolor=blue,linkcolor=blue]{hyperref}
\usepackage{datetime}
\usepackage{longtable}
\usepackage{natbib}
\usepackage{amsmath}
\usepackage{ulem}
\usepackage{bm}
\usepackage[usenames,dvipsnames]{xcolor}

\usepackage{etoolbox}

\makeatletter

\patchcmd{\NAT@citex}
  {\@citea\NAT@hyper@{%
     \NAT@nmfmt{\NAT@nm}%
     \hyper@natlinkbreak{\NAT@aysep\NAT@spacechar}{\@citeb\@extra@b@citeb}%
     \NAT@date}}
  {\@citea\NAT@nmfmt{\NAT@nm}%
   \NAT@aysep\NAT@spacechar\NAT@hyper@{\NAT@date}}{}{}

\patchcmd{\NAT@citex}
  {\@citea\NAT@hyper@{%
     \NAT@nmfmt{\NAT@nm}%
     \hyper@natlinkbreak{\NAT@spacechar\NAT@@open\if*#1*\else#1\NAT@spacechar\fi}%
       {\@citeb\@extra@b@citeb}%
     \NAT@date}}
  {\@citea\NAT@nmfmt{\NAT@nm}%
   \NAT@spacechar\NAT@@open\if*#1*\else#1\NAT@spacechar\fi\NAT@hyper@{\NAT@date}}
  {}{}

\makeatother

\newcommand{\noprint}[1]{}

\usepackage{color}

\newcommand{\lsun}{{L$_\odot$}}
\newcommand{\msun}{{M$_\odot$}}
\newcommand{\mjup}{{M$_\textrm{Jup}$}}

\newcommand{\thisstar}{GJ\,3305}

\begin{document}
\title{Dynamical Masses of Young M Dwarfs: Masses and Orbital Parameters
of \thisstar\,AB, the Wide Binary Companion 
to the Imaged Exoplanet Host 51\,Eri$^{*}$}

\newcommand{\caltech}{1}
\newcommand{\cfa}{2}
\newcommand{\jcpa}{3}
\newcommand{\asu}{4}
\newcommand{\lowell}{5}
\newcommand{\yale}{6}
\newcommand{\scsu}{7}
\newcommand{\hawaii}{8}
\newcommand{\texas}{9}

\author{%
Benjamin~T.~Montet\altaffilmark{\caltech,\cfa},
Brendan~P.~Bowler\altaffilmark{\caltech,\jcpa},
Evgenya~L.~Shkolnik\altaffilmark{\asu, \lowell},
Katherine~M.~Deck\altaffilmark{\caltech,\jcpa},
Ji~Wang\altaffilmark{\caltech, \yale},
Elliott~P.~Horch\altaffilmark{\scsu},
Michael~C.~Liu\altaffilmark{\hawaii},
Lynne~A.~Hillenbrand\altaffilmark{\caltech},
Adam~L.~Kraus\altaffilmark{\texas},
David~Charbonneau\altaffilmark{\cfa}
}

\email{btm@astro.caltech.edu}
\altaffiltext{\caltech}  {Cahill Center for Astronomy and Astrophysics,
                          California Institute of Technology, Pasadena, CA,
                          91125, USA}
\altaffiltext{\cfa}      {Harvard-Smithsonian Center for Astrophysics,
                          Cambridge, MA 02138, USA}
\altaffiltext{\jcpa} {Caltech Joint Center for Planetary Astronomy Fellow}
\altaffiltext{\asu} {School of Earth and Space Exploration, Arizona State University, 
                            Tempe, AZ 85287, USA}
\altaffiltext{\lowell}   {Lowell Observatory, 1400 West Mars Hill Road, 
                          Flagstaff, AZ, 86001, USA}
\altaffiltext{\yale}{Department of Astronomy, Yale University, 
                        New Haven, CT 06511, USA}
\altaffiltext{\scsu}     {Department of Physics, Southern Connecticut State University, 
                         501 Crescent Street, New Haven, CT 06515, USA}
\altaffiltext{\hawaii}   {Institute for Astronomy, University of Hawaii, 
                          2680 Woodlawn Drive, Honolulu, HI 96822, USA}
\altaffiltext{\texas} {Department of Astronomy, The University of Texas at
                            Austin, Austin, TX 78712, USA}
\altaffiltext{*}{Some of the data presented herein were obtained at the W.M. Keck Observatory, which is operated as a scientific partnership 
among the California Institute of Technology, the University of California and the National Aeronautics and Space Administration. 
The Observatory was made possible by the generous financial support of the W.M. Keck Foundation.}

\date{\today, \currenttime}

\begin{abstract}
We combine new high resolution 
imaging and spectroscopy from Keck/NIRC2, Discovery Channel Telescope/DSSI, and Keck/HIRES 
with published astrometry and radial velocities to measure individual 
masses and orbital elements of the \thisstar\,AB system, a young ($\sim$20 Myr) M+M binary 
(unresolved spectral type M0) member of the $\beta$ Pictoris moving group
comoving with the imaged exoplanet host 51\,Eri. We measure a total system mass of $1.11 \pm 0.04$ \msun, a 
period of $29.03 \pm 0.50$ year, a semimajor axis of $9.78 \pm 0.14$ AU, and
an eccentricity of $0.19 \pm 0.02$.
The primary component has a dynamical mass of $0.67 \pm 0.05$ \msun\ and the 
secondary has a mass of $0.44 \pm 0.05$ \msun. 
The recently updated BHAC15 models are consistent
with the masses of both stars to within $1.5\sigma$.
Given the observed masses the models predict an age of the \thisstar\,AB system of $37 \pm 9$ Myr.
Based on the the observed system architecture and our dynamical mass measurement, 
it is unlikely that the orbit of 51\,Eri\,b has been significantly altered by the Kozai-Lidov mechanism.
\end{abstract}


\maketitle

\section{Introduction}
\label{sec:intro}

Loose associations
of young, nearby ($<$70 pc) stars with common ages,
kinematics, and origins have been a subject of increasing interest
\citep{Zuckerman04, Shkolnik12, Malo13}.
Because of their proximity to Earth, these young
moving groups (YMGs) are excellent targets to study pre-main sequence (PMS)
stellar and substellar evolution, protoplanetary and debris disk structure,
and giant planet formation at ages between distant
star-forming regions and old field stars \citep[e.g.][]{Close05, Nielsen10}. 
About 10 YMGs containing hundreds of objects between 8 and 120 million years old 
are known \citep[e.g.][]{Torres08}.

As these moving groups are amenable to numerous age dating methods, including kinematic techniques, 
they provide the opportunity 
to measure dynamical masses of PMS low-mass binary 
objects and test stellar evolution models \citep{Stassun14}.
Generally, PMS stellar
masses are inferred by comparing a star's temperature,
luminosity and metallicity to model predictions
\citep[e.g.][]{Schaefer14}. 
These models
are poorly constrained by observations and may induce systematic
offsets \citep{Dupuy09, Dupuy14}. 
Worse yet, different models predict disparate masses, primarily
due to uncertainties in the treatment of convection in
low-gravity atmospheres \citep{Baraffe02}, stellar
accretion history \citep{Baraffe10}, and molecular line lists
\citep{Baraffe15}.
In some cases, model-predicted masses can differ by a factor of two or
more \citep{Hillenbrand04, Schlieder14}. 
Dynamical mass measurements of binary stars with known ages 
are essential to test models.

Recently, \citet{Macintosh15} presented 51\,Eri\,b, the first
exoplanet discovery from the Gemini Planet Imager.
The planet has a mass of $\approx 2$ \mjup\ (assuming a hot start model), 
a projected separation of 13 AU,
a temperature of 600-750 K, and a T4.5-T6 spectral type.
\thisstar\ is known to be a binary with combined spectral type M0 
\citep{Kasper07}.
\citet{Feigelson06} identified \thisstar\ and 51\,Eri as an F0-M0 common proper motion pair, 
separated by $66 \arcsec$ or $\sim$2000 AU.

As a binary system,
a dynamical mass can be measured for both stars in \thisstar\,AB. 
As both stars are members of the $\beta$ Pictoris moving group, an approximate age of
the system is known \citep[$24 \pm 3$ Myr;][]{Binks14, Mamajek14, Bell15}.
While most dynamical masses of M dwarfs are limited by distance uncertainties, 
51\,Eri has a parallax from \textit{Hipparcos} measured
to a precision of 1\%.
Combining this parallax with 15 years of imaging and RV data enables us to determine
the system orbital parameters, elucidating
the architecture of this 4---or more---body system.

In this paper, we combine RV and astrometric observations of \thisstar\,AB 
to measure orbital parameters and masses for each star. 
We compare these masses to model predictions and
discuss the possible implications of this binary pair on the long-term evolution of
the orbit of 51\,Eri\,b.

\section{Data Collection and Reduction}
\label{sec:data}

\thisstar\,AB has been imaged and resolved many times
\citep{Kasper07, Bergfors10, Delorme12, Janson12, Janson14a}. 
The system was also imaged with NIRC2 \citep{Wizinowich00} in one unpublished epoch
in 2001 available in the Keck Observatory Archives (KOA, PI Zuckerman). 
In this work, we combine these data with five observations from 2002 to 2015, three
using Keck/NIRC2 and one with the Differential Speckle Survey Instrument \citep[DSSI,][]{Horch09} at 
the Discovery Channel Telescope at Lowell Observatory.

All NIRC2 data were obtained with the narrow camera mode, which has a field of view 
of 10$\farcs$2 $\times$ 10$\farcs$2 and a plate scale of 9.952~mas pixel$^{-1}$
\citep{Yelda10}.  All images were flat fielded and cleaned of bad pixels and 
cosmic rays.  Astrometry and relative photometry of \thisstar\ was derived by 
simultaneously fitting three bivariate Gaussians to each component following 
\citet{Liu10}. 

DSSI allows for simultaneous observations in two filters.
We use the DSSI $R$ and $I$ filters, with central wavelengths
692 and 880~nm and FWHMs of 40 and 50~nm. 
We obtained 1000 40-ms exposures in each channel simultaneously.
The data were then reduced following \citet{Horch15}.
Specifically, the autocorrelation of each frame was calculated and summed over all
exposures, and the near-axis subplanes of the image bispectrum were 
calculated. 
To create a reconstructed image, the Fourier transform of the autocorrelation
of the binary was divided by that of a nearby point source (HR\,1415).
The square root of this value is taken, and the result combined with a phase function
derived from the bispectral subplanes.
The pixel scale (19~mas pixel$^{-1}$ in $R$ and 20~mas pixel$^{-1}$ in $I$) and orientation of the detector were found by observing
several widely separated binaries with known astrometry.
Our astrometry is listed in Table~\ref{tab:astrometry}.

\begin{deluxetable*}{lcccccc}
\tablecaption{Data for \thisstar\,AB}
\footnotesize
\tablewidth{0pt}
\tablehead{
  \colhead{Epoch} & 
  \colhead{Bandpass} &
  \colhead{RV} &
  \colhead{Contrast} &
  \colhead{Separation}     &
  \colhead{Position Angle} &
  \colhead{Source} \\
  \colhead{(Year)} &
  \colhead{} &
  \colhead{(km s$^{-1}$)} &
  \colhead{($\Delta$ mag)} &
  \colhead{(mas)} &
  \colhead{(deg)} &
  \colhead{}
}
\startdata
2001.910 & H$_2$($\nu$=1--0) & & $1.00 \pm 0.02$ & $286  \pm 1$   & $198.1 \pm 0.1$   & This Work \\
2002.162 & $H$               & & $1.02 \pm 0.02$ & $275.4\pm 1.5$ & $197.9 \pm 0.2$   & This Work \\
2003.05  & $K$              &  & $0.94 \pm 0.05$ & $225  \pm 5\altaffilmark{1}$   & $195.0 \pm 1.5\altaffilmark{1}$   & \citet{Kasper07} \\
2003.195 & $H$              &  & $0.99 \pm 0.01$ & $217  \pm 1$   & $196.8 \pm 0.1$   & This Work \\
2004.02  & $L'$              & & & $159  \pm 2$   & $194   \pm 1$     & \citet{Delorme12} \\
2004.95  & $L'$              & & $0.88 \pm 0.28$ & $93   \pm 2$   & $189.5 \pm 0.4$   & \citet{Kasper07} \\
2008.88  & SDSS $z'$       &   & $1.39\pm 0.16$ & $218  \pm 2$   & $20.3  \pm 0.3$   & \citet{Bergfors10} \\
2008.88  & SDSS $i'$        &  & $2.57\pm 0.05$ & $218  \pm 2$   & $20.3  \pm 0.3$   & \citet{Bergfors10} \\
2009.13  & SDSS $i'$+$z'$    & &                & $231  \pm 2$   & $19.2  \pm 0.3$   & \citet{Janson12} \\
2009.90  &  $L'$             &   &              & $269  \pm 3$   & $18.6  \pm 1.0$   & \citet{Delorme12} \\
2009.98  &  $L'$             &    &             & $272  \pm 3$   & $19.2  \pm 1.0$   & \citet{Delorme12} \\
2010.10  & SDSS $z'$       &   & $1.34\pm 0.01$ & $284  \pm 3$   & $18.5  \pm 0.6$   & \citet{Janson12} \\
2010.10  & SDSS $i'$       &   & $3.73\pm 0.01$ &                &                   & \citet{Janson12} \\
2010.81  & SDSS $z'$       &   &                & $297  \pm 3$   & $19.4  \pm 0.3$   & \citet{Janson14a} \\
2011.67  &  $L'$             &    &             & $303  \pm 3$   & $18.1  \pm 1.0$   & \citet{Delorme12} \\
2011.87  & SDSS $z'$        &  &                 & $295  \pm 4$   & $18.5  \pm 0.3$   & \citet{Janson14a} \\
2012.01  & SDSS $z'$        &  &                 & $307  \pm 3$   & $18.2  \pm 0.3$   & \citet{Janson14a} \\
2014.629 & Br$\gamma$    &     & $0.92 \pm 0.01$ & $244  \pm 1$   & $16.8 \pm 0.1$    & This Work \\
2014.746 & DSSI $R$       &    & $1.89 \pm 0.04$ & $239\pm 1$ & $16.4 \pm 0.2$  & This Work \\
2014.746 & DSSI $I$        &   & $1.17 \pm 0.03$ &    $240 \pm 1$       &     $16.1 \pm 0.2$           & This Work\\
2015.653 & $K$               &    &  $0.93 \pm 0.01$ &    $199 \pm 1$      &   $15.6 \pm 0.1$     & This Work \\
2015.653 & $H$               &    &  $0.99 \pm 0.01$ &   $198 \pm 1$    &  $15.6 \pm 0.1$        & This Work \\
2015.653 & $J$               &    &  $0.97 \pm 0.01$ &    $199 \pm 1$   &  $15.6 \pm 0.2$      & This Work \\
2015.653 & $Y$               &    &  $1.06 \pm 0.03$ &   $200 \pm 1$    &  $15.6 \pm 0.1$      & This Work \\
\hline
2003.796   & HIRES $V$   & $ 19.41 \pm 0.38 $ &        &         &          & This work \\
2004.884 & NIRSPEC $K$ &  $19.86 \pm 0.05$         &        &       &         &  \citet{Bailey12} \\
2005.862 & NIRSPEC $K$ &  $20.55 \pm 0.06$         &        &       &         &  \citet{Bailey12} \\
2005.971 & HIRES $V$ &      $21.70 \pm 0.30$     &        &       &         &  \citet{Shkolnik12} \\
2006.014 & NIRSPEC $K$ &  $20.82 \pm 0.05$         &        &       &         &  \citet{Bailey12} \\
2006.016 & NIRSPEC $K$ &  $20.95 \pm 0.05$         &        &       &         &  \citet{Bailey12} \\
2006.019 & NIRSPEC $K$ &  $20.95 \pm 0.05$         &        &       &         &  \citet{Bailey12} \\
2011.778 & UVES Blue &        $24.40 \pm 0.04$   &        &       &         &  \citet{Elliott14} \\
2001.994 & UVES Blue &        $23.30 \pm 0.02$   &        &       &         &  \citet{Elliott14} \\
2012.022 & UVES Blue &        $23.80 \pm 0.02$   &        &       &         &  \citet{Elliott14} 
\enddata
\tablenotetext{1}{Observations published without uncertainty estimates; we choose conservative
values.}
\tablecomments{In some previous analyses, contrast ratios were not listed for specific epochs.
Observations without listed separations correspond to simultaneous multiband photometry.}
\label{tab:astrometry}
\end{deluxetable*}

The \thisstar\ binary system has also been monitored spectroscopically. 
One Keck/HIRES spectrum from 2003 exists in the KOA (PI Zuckerman); we measure
the RV following \citet{Kraus15}.
We combine this spectrum with nine additional spectra from \citet{Bailey12}, \citet{Shkolnik12},
and \citet{Elliott14}.
In all cases, the RVs were calculated treating the system as an SB1.
We take the reported RV and uncertainty for each observation, but assume the flux from the secondary
is non-negligible, as explained in Section~\ref{sec:analysis}.

\section{Analysis}
\label{sec:analysis}

We infer the orbital parameters of \thisstar\,AB by comparing the astrometric
and RV data to a Keplerian orbit model at each of the observation
times.
A parallax, astrometric orbit, and SB1 RV data can be combined to measure individual
masses of each star \citep[e.g.][]{Bean07}.
There is no measured parallax for \thisstar, so we adopt the Hipparcos distance to 51\,Eri/HIP\,21547: 
$29.43 \pm 0.30$ pc \citep{vanLeeuwen07}.
These two comoving systems have a projected separation of $1940 \pm 20$ AU,
or 0.01 pc. It is unlikely that the radial distance between the two could
be significantly larger while remaining bound; we apply this parallax
as a prior on the distance to \thisstar.

We then fit for nine additional parameters that define the orbits of
the two stars as viewed from Earth.
Of these, seven can be obtained from astrometry.
These parameters are the eccentricity vectors $\sqrt{e} \cos{\omega}$ and
$\sqrt{e} \sin{\omega}$, the time of periapse $t_P$, the period $P$, 
the total mass $M_1 + M_2$, the inclination $i$, and the 
longitude of the ascending node $\Omega$.
We parameterize the eccentricity vector in this manner following \citet{Eastman13}.
The RV data can provide additional information about several of these
(not $M_1 + M_2$ or $\Omega$ directly), also allowing us to fit the
systemic RV $\gamma$ and the secondary mass $M_2$.

We include ten additional terms to account for 
possible systematics in the datas. 
This star has been imaged, resolved, and published by four different groups.
We account for the possibility each group may have underestimated 
their uncertainties on the orbital separation and position angle
by a multiplicative factor by including a systematic error term on the measured positions 
from each group, allowing outlier points to be downweighted without manually choosing
specific points to downweight.
We do the same with our reductions of both archival and new data,
allowing for separate systematic error terms on our data from Keck/NIRC2 and 
DCT/DSSI, providing a total of six systematic error terms.
We allow the uncertainties on each dataset to be inflated up
to a factor of five.

Similarly, we allow for the possibility that the uncertainties in the 
RVs may be underestimated, possibly due to stellar variability \citep{Moulds13},
errors in systemic RVs of standard stars, or drifts in the stability of the
spectrographs.
As our RV data originate from four sources, we allow each to have its own 
systematic error term, analogous to the jitter term commonly applied 
in RV orbit fits of exoplanets \citep[e.g.][]{Johnson11b}:
\begin{equation}
\log \mathcal L \propto - \sum_i \bigg[\log{\sqrt{\sigma_{o,i}^2 + \sigma_s^2}} + 0.5 \bigg(\frac{(f_i(t) -
v_i(t))^2}{\sigma_{o,i}^2 + \sigma_s^2}\bigg)\bigg].
\end{equation}
Here, $\mathcal L$ is the likelihood of the data given some underlying physical model,
$\sigma_{o,i}$ is the observed uncertainty on the $i$th data point, $\sigma_s$ 
the systematic error associated with each particular set of observations, $f_i(t)$ the RV
model evaluated at time $t$, and $v_i(t)$ the observed RV at each $t$. 
Maximum likelihood jitter values range from $0.13$ km s$^{-1}$ for the 2003 HIRES data
to $0.57$ km s$^{-1}$ for the UVES data, suggesting stellar jitter is significant
in the RV data, as expected for young stars. 

In all cases, one set of lines are observed because the RV separation is smaller than
the line width.
We expect each RV measurement to be the flux-weighted sum of the two individual
RVs.
At each step, we calculate the RVs for each star, weighting them according to their 
expected flux contribution in each bandpass, using the observed flux ratios in the
visible and near-IR as priors and assuming an additional 0.1~mag of variability in 
the optical and 0.05~mag in the near-IR.

We neglect the possibility that 51\,Eri could contribute significantly to the observed RV signal. 
Following Equation~1 of \citet{Montet15a}, 
the maximum RV acceleration expected from 51\,Eri is 3 cm s$^{-1}$ yr$^{-1}$,
well below our sensitivity.

\begin{figure*}[htbp!]
\centerline{\includegraphics[width=0.85\textwidth]{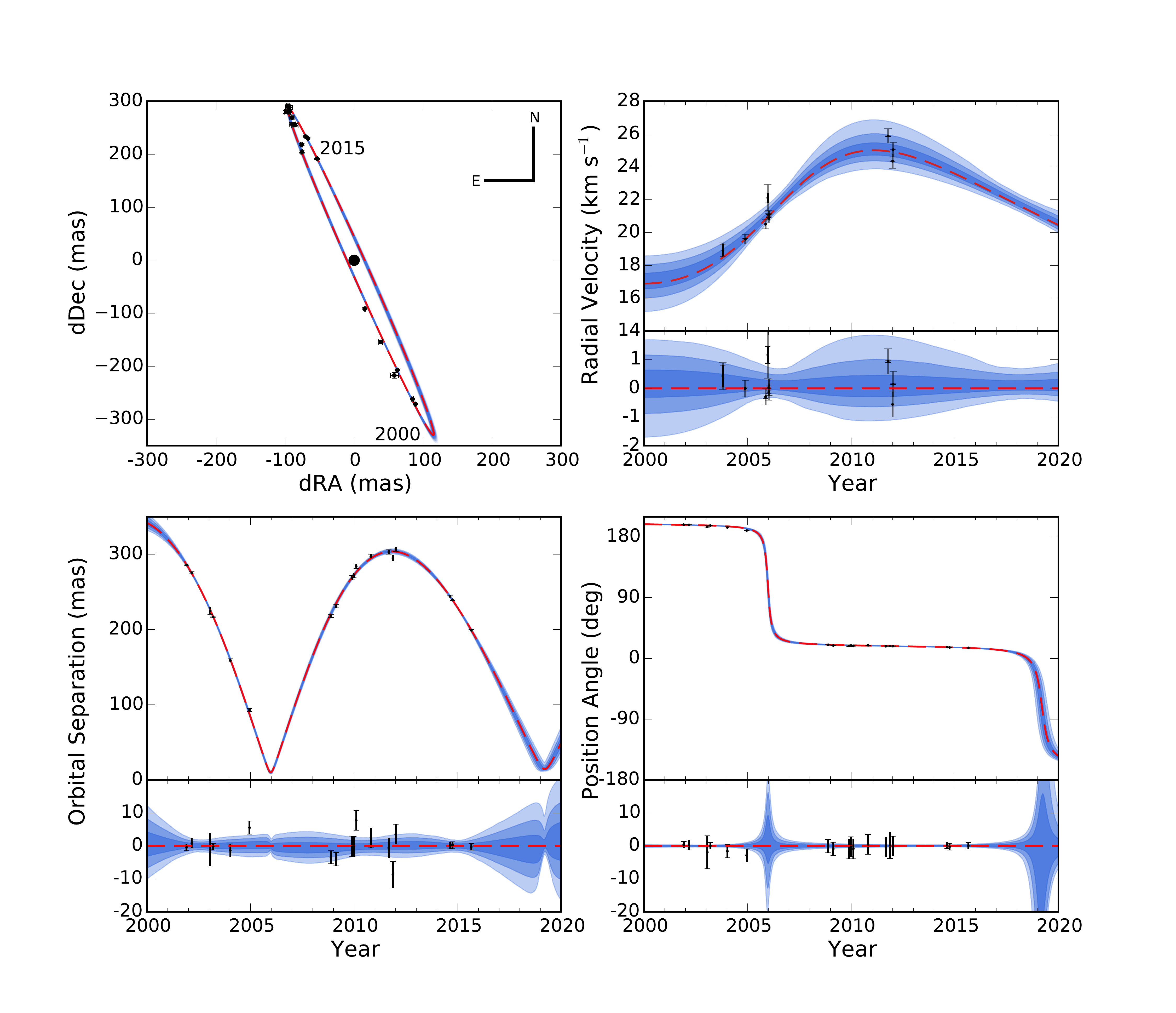}}
\caption{(Top Left) Astrometry for \thisstar\,AB. Data points correspond to the observations
listed in Table \ref{tab:astrometry}. Blue lines correspond to random draws from the posterior
distributions of orbital elements. The red, dashed line corresponds to the maximum likelihood
orbit. (Top Right) RV data for \thisstar\,A from the literature. 
The published uncertainties are in black;
in gray are the best-fitting uncertainties, incorporating an RV jitter model. The red, dashed
line corresponds to the maximum likelihood orbit. The blue shaded regions correspond
to the $1-$, $2-$, and $3\sigma$ uncertainties in the RV of \thisstar\,A.
(Bottom Left) Measured separations for \thisstar\,AB and residuals from the maximum likelihood
model. 
Each feature on the plot retains its meaning from the previous subplot.
(Bottom Right) Measured position angles for \thisstar\,AB and residuals from the maximum 
likelihood model.
  }
\label{fig:fits}
\end{figure*}

We calculate posterior distributions for all parameters using
\texttt{emcee} \citep{Foreman-Mackey12}, an implementation of the
affine-invariant Markov Chain Monte Carlo ensemble sampler of \citet{Goodman10}.
After performing a local optimization to determine a maximum-likelihood fit, 
we move 3000 walkers each 4000 steps.
We discard the first 2000 steps of each walker as burn-in, and use the test of
\citet{Geweke92} and visual inspection to verify the 
system has converged.
The data and allowed orbits are shown in Figure~\ref{fig:fits}.
Summary statistics for the orbital parameters are given in
Table~\ref{tab:results}.
We note the fitted systemic RV of $20.76 \pm 0.18$ km s$^{-1}$ is consistent with the 
measured RV for 
51\,Eri, $21.0 \pm 1.2$ km s$^{-1}$ \citep{Bobylev06} and the UVW velocities
are consistent with \citet{Mamajek14}.
Our samples are available 
online.\footnote{http://www.astro.caltech.edu/$\sim$btm/research/gj3305.html}

\begin{deluxetable}{lccc}
\tablecaption{Parameters for \thisstar\,AB}
\footnotesize
\tablewidth{0pt}
\tablehead{
  \colhead{Parameter} & 
  \colhead{Median}     &
  \colhead{} &
  \colhead{Uncertainty}  \\
  \colhead{} & 
  \colhead{} &
  \colhead{} &
  \colhead{($1\sigma$)}      
}
\startdata
$\sqrt{e}\cos{\omega}$ & 0.160 & $\pm$ & 0.019  \\
$\sqrt{e}\sin{\omega}$ & -0.406 & $\pm$ & 0.015 \\
Eccentricity & 0.19 & $\pm$ & 0.02 \\ 
Argument of Periastron $\omega$~[deg] & -69 & $\pm$ & 3 \\  
Time of Periastron [Year] & 2007.14 & $\pm$ & 0.16 \\
Orbital Period [Year] & 29.03 & $\pm$ & 0.50 \\
\thisstar\,A Mass [\msun] & 0.67 & $\pm$ & 0.05 \\
\thisstar\,B Mass [\msun] & 0.44 & $\pm$ & 0.05 \\  
Total System Mass [\msun] & 1.11 & $\pm$ & 0.04 \\
Mass Ratio $M_B/M_A$      & 0.65 & $\pm$ & 0.10 \\  
Orbital Inclination, $i$~[deg] & 92.1 & $\pm$ & 0.2 \\  
Orbital Semimajor Axis, $a$~[AU] & 9.78 & $\pm$ & 0.14 \\ 
Long. of Ascending Node, $\Omega$~[deg] & 18.8 & $\pm$ & 0.2 \\ 
Systemic RV Velocity, $\gamma$~[km s$^{-1}$] & 20.76 & $\pm$ & 0.18 \\
RV semiamplitude $K_A$~[km s$^{-1}$] & 4.01 & $\pm$ & 0.38  \\ 
U [km s$^{-1}$] & -13.76 & $\pm$ & 0.24 \\
V [km s$^{-1}$] & -16.40 & $\pm$ & 0.40 \\
W [km s$^{-1}$] & -9.71  & $\pm$ & 0.36 \\
\thisstar\,A Luminosity [\lsun] & 0.112 & $\pm$ & 0.007 \\  
\thisstar\,B Luminosity [\lsun] & 0.043 & $\pm$ & 0.005 
\enddata
\label{tab:results}
\end{deluxetable}

We estimate bolometric luminosities for both stars by integrating the CFHIST2011\_2015 model 
spectra of \citet{Baraffe15}.
We use the 3700 and 3500 K models with $\log g = 4.5$ (cgs) as spectral templates, scaling them
until they match the observed combined and differential magnitudes in each available bandpass.
We add in quadrature 0.10~mag of uncertainty in our visible-light magnitudes and 0.05~mag in 
the near-IR to account for stellar variability.

\section{Comparison with BHAC15 Evolutionary Models}
\label{sec:models}

Given the known distance to the system from \textit{Hipparcos} 
we can test if theoretical stellar evolution models
accurately predict the inferred stellar masses and age of the $\beta$ Pic moving group.
Combined-light photometry spanning from $B$ (0.4 $\mu$m) to $Ks$ (2.3 $\mu$m) was measured by the 
APASS, 2MASS, and \textit{WISE} surveys (Table~\ref{tab:photometry}).
We add an uncertainty of 0.03 mag in quadrature to the listed APASS uncertainties due to
the presence of systematics in APASS DR9 at that level \citep{Henden12}.
We also have obtained one epoch of differential photometry in two visible-light bandpasses with
DSSI and two near-IR bandpasses ($H$ and Br$\gamma$) with Keck/NIRC2.

\begin{deluxetable}{lccc}
\tablecaption{Photometry for \thisstar\,AB}
\footnotesize
\tablewidth{0pt}
\tablehead{
  \colhead{Bandpass} & 
  \colhead{Source}     &
  \colhead{Magnitude} &
  \colhead{Uncertainty}
}
\startdata
Combined & & & \\
$B$ &  APASS DR9 & 11.94 & 0.03  \\
$V$ &  APASS DR9 & 10.56 & 0.05 \\
$g'$ & APASS DR9 & 11.27 & 0.03 \\
$r'$ & APASS DR9 & 10.03 & 0.07 \\
$J$ &  2MASS & 7.30 & 0.02 \\
$H$ &  2MASS & 6.64 & 0.05 \\
$K$ &  2MASS & 6.41 & 0.02 \\
$W1$ & WISE & 6.34 & 0.03 \\
$W2$ & WISE & 6.21 & 0.02 \\
$W3$ & WISE & 6.16 & 0.02 \\
$W4$ & WISE & 6.00 & 0.04 \\
\hline
Resolved & & & \\
$\Delta$692 & DSSI & 1.89 & 0.04 \\
$\Delta$880 & DSSI & 1.17 & 0.03 \\
$\Delta$H$_2$ & Keck/NIRC2 & 1.00 & 0.02 \\
$\Delta$Br$\gamma$ & Keck/NIRC2 & 0.92 & 0.01 \\
$\Delta H$ & Keck/NIRC2 & 1.00 & 0.02
\enddata

\label{tab:photometry}
\end{deluxetable}

We compare the observed brightness of \thisstar\,AB to that predicted by the BHAC15 models
of \citet{Baraffe15} for two stars of masses consistent with those inferred during our analysis
as a function of age. 
We find models of 25 Myr old stars accurately predict the combined-light near-IR flux for these
stars, although the models predict brighter $V$ magnitudes than those observed (Figure 2).
However, star $B$ is brighter than these same models predict: a 25 Myr old \thisstar\,B would be 
significantly brighter than what is observed. 
Assuming the stars are coeval, the models then predict a mass for \thisstar\,B that is 20\% lower 
than the observed mass. 

We create a simulated spectral energy distribution for each star, given the measured masses and the average age of 
$\beta$ Pic as measured from higher-mass stars. 
We interpolate absolute magnitudes predicted by the updated BHAC15 models of \citet{Baraffe15} along 
isochrones and isomass contours to predict apparent 
magnitudes for these stars in each bandpass.
We find that the total received flux is lower than
predicted by the BHAC15 models in each bandpass. While the flux for \thisstar\,A is consistent with 
the model predictions, \thisstar\,B is fainter than predicted.

Given the observed masses, we then vary the age of the system, assuming both stars are coeval, 
to determine which system age would be predicted by
these models given the observed combined and differential magnitudes.
We apply a flat prior on the age of the system, finding the BHAC15 models predict an age of $37 \pm 9$ Myr, consistent with the overall
age of the moving group \citep[$24 \pm 3$ Myr,][]{Bell15}.
As the system is unambiguously young, we can also confirm 51\,Eri\,b as 
a planetary mass object.

\begin{figure*}[htbp!]
\centerline{\includegraphics[width=0.82\textwidth]{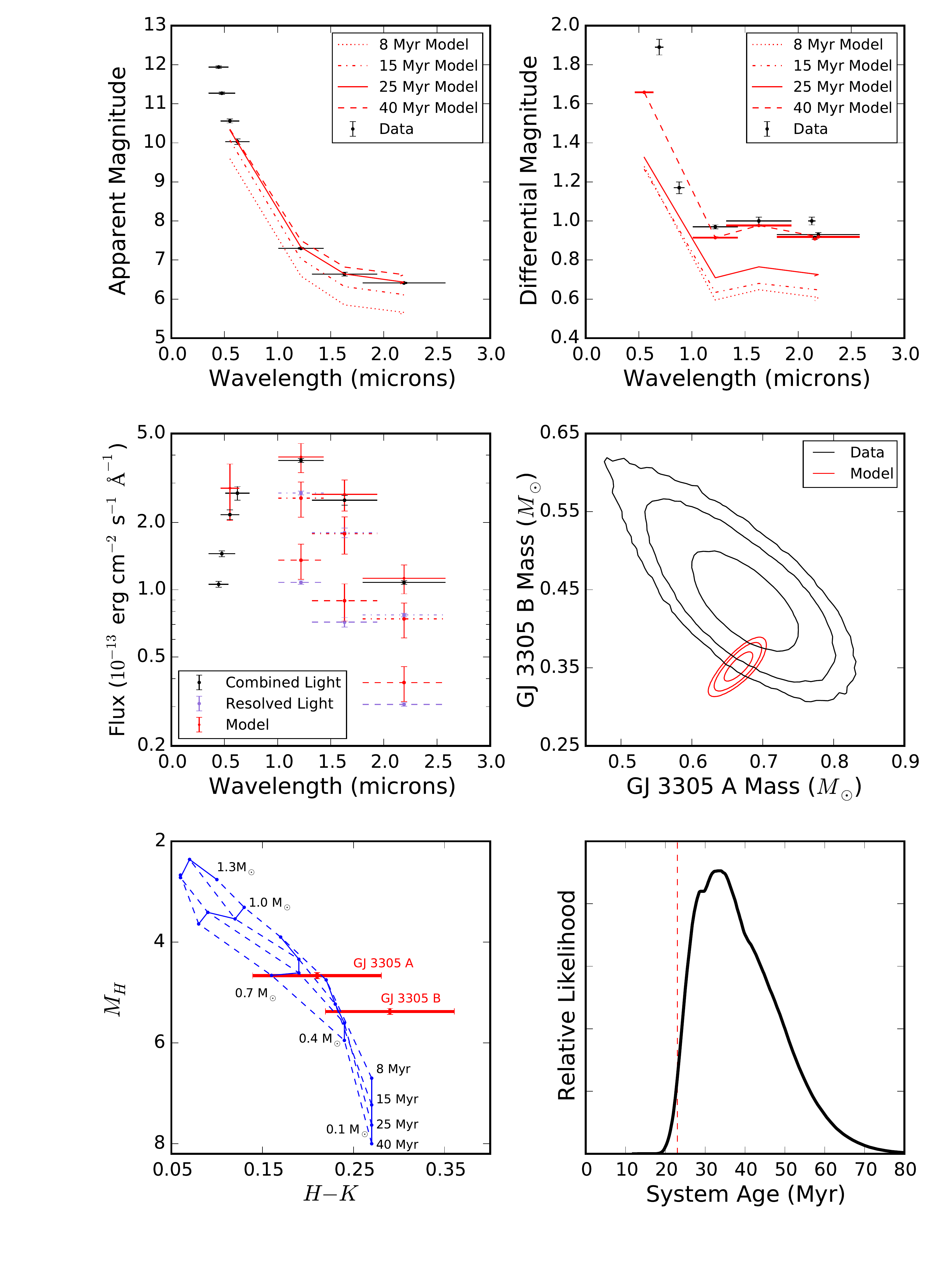}}
\caption{(Top) (Left) Combined-light, unresolved and (Right) differential, resolved photometry for \thisstar\,AB (black) compared to
predictions (red) of the BHAC15 models as a function of age given the observed masses and parallax. The data are consistent with an age larger than 25 Myr.
Plotted bars along the abscissa correspond to the width of each filter and are meant to guide the eye: 
they do not represent an uncertainty.
(Middle left) SED for the system, assuming a $24 \pm 3$ Myr age and the observed masses. 
Combined-light photometry is in black and resolved photometry in purple. 
While the
model accurately reproduces the observed flux from \thisstar\,A, it overpredicts the received flux from 
\thisstar\,B.
(Middle right) Joint posterior probability distributions on the masses of the two stars,
(black) inferred from the astrometry and RV data and (red) predicted by the BHAC15 
models given the observed combined-light and differential photometry assuming
an age of $24 \pm 3$ Myr. Contours correspond
to the 1-, 2-, and 3-$\sigma$ confidence regions. The BHAC15 models predict a mass for
\thisstar\,B consistent with the mass inferred from the data, but underpredicts
the mass of \thisstar\,A by 20\%.
(Bottom left) CMD showing the absolute $H$ magnitudes and $H-K$ colors of \thisstar\,AB compared to theoretical
models. The models provide a more accurate fit for \thisstar\,A than \thisstar\,B.  
(Bottom right) Posterior probability distribution on the age of the \thisstar\ system, calculated by marginalizing the joint mass-age posterior over
all allowed masses, assuming both stars are the same age.
The BHAC15 models predict an age of $37 \pm 9$ Myr; the dashed line represents the \citet{Bell15} age of the $\beta$ Pictoris system.}
\label{fig:models}
\end{figure*}

\section{Discussion}

We have measured the masses and orbits of \thisstar\,AB, finding both to be 
consistent with the BHAC15 models at the $1.5\sigma$ level. 
In the future \thisstar\,AB and the gravitationally bound 51\,Eri\,Ab will be able to act as an isochronal test as a coeval,
co-metallicity quadruple system spanning stellar to planetary mass regimes.

The derived period of \thisstar\ ($29.03 \pm 0.50$ year) is longer than the 21 year found by \citet{Delorme12}.
The authors of that paper did not have sufficient data to fit all orbital parameters,
so they fixed the total system mass to $1.3$~\msun. Given our lower mass measurement, 
it is not surprising 
that our measured orbital period is longer. 

\subsection{Current Limitations}

It is possible that an unseen very low-mass star or brown dwarf orbiting
\thisstar\,B could cause us to overestimate its mass, causing the observed
$20\%$
discrepancy. 
For the system to be stable over $20$ Myr, such a companion would have to
be in a close ($P < 50$ day) orbit. 
The companion would then have to be in a nearly face-on ($i < 10^{\circ}$) orbit to evade RV detection.
Such companions could be found through continued astrometric monitoring of \thisstar. 
Such a companion would not affect our astrometry due to its small separation from \thisstar\,B
and would likely not affect our photometry due to its low luminosity relative to the other stars
in the system.

Most PMS M dwarfs have distance measurements to a precision no better than 5\%, meaning the 
total mass cannot be measured to better than 15\% \citep[e.g.][]{Shkolnik12}.
The uncertainty in the mass of \thisstar\,AB is only 4\%: the dominant source
of uncertainty in this value is the 1\% \textit{Hipparcos} parallax
to 51\,Eri,
making this system an ideal low-mass benchmark.
With a Gaia parallax forthcoming in the next few years, parallaxes for low-mass PMS stars will
be improved substantially.
Long-term astrometric and RV monitoring of wide M dwarfs is essential
as parallaxes are obtained over the next few years.

The uncertainty in the individual mass of each star is dominated by the uncertainty in the 
Doppler semiamplitude. 
While additional astrometric observations will not significantly improve the measured 
physical properties of \thisstar, additional RV observations will be important.
RV observations behind AO would be especially beneficial, as the RV from each 
star could be measured separately, instead of a flux-weighted RV centroid.

\subsection{Dynamical Effects on 51\,Eri\,b}

GJ 3305 AB and 51\,Eri\,Ab exist in a
dynamical configuration that may be susceptible to Kozai-Lidov
oscillations
\citep{Kozai62,Lidov62}, as suggested by \citet{Macintosh15}. 
In this scenario, the
planet-star binary (51\,Eri\,Ab) interacts secularly with \thisstar\,AB, 
leading to oscillations in inclination and eccentricity of the planet-star
sub-system. The timescale for such an interaction is
\begin{equation}
\tau \approx
P_\textrm{planet}\frac{M_\star}{M_\textrm{pert}}\bigg(\frac{a_\textrm{pert}}{a_\textrm{planet}}\bigg)^3(1-e_\textrm{pert}^2)^{3/2}
\end{equation}
where $P_{planet}$ is the orbital period of a planet with a semimajor
axis of $a_{planet}$ about a host of mass $M_\star$,
$M_\textrm{pert}$ is
the mass of a distant perturber, and $a_\textrm{pert}$ and $e_\textrm{pert}$
are the semimajor axis
and eccentricity of the perturber/planet-star ``binary'' orbit 
\citep[see e.g.][]{Holman97}.

Although we have limited information about this system, we can
estimate the timescale for Kozai-Lidov cycles should the mutual
inclination of the 51\,Eri\,Ab system and (51\,Eri\,Ab)-(\thisstar\,AB)
system satisfy $ 140^\circ \lesssim i_m  \gtrsim 40^\circ$. 
Taking the instantaneous sky-projected separations as a
proxy for the semimajor axes and inferred masses of
$M_\star=1.75$~\msun\  \citep{Simon11} and $M_\textrm{pert} = 1.1$~\msun\ yields a
timescale of
$\tau \sim 2\times10^8 \textrm{ yr }(1-e_\textrm{pert}^2)^{3/2}$.
Therefore, unless the eccentricity
of \thisstar\ about the 51\,Eri subsystem satisfies
$e_\textrm{pert}\gtrsim 0.9$, the timescale for Kozai-Lidov oscillations
is longer than the age of the system,
so we do not expect the Kozai-Lidov mechanism to have had
time to induce a large eccentricity or spin-orbit misalignment
within the 51\,Eri sub-system. If future
observations indicate non-zero spin-orbit misalignment or a high
eccentricity for the orbit of 51\,Eri\,b, a primordial origin unrelated to
the distant perturbers would be suggested.

\acknowledgements

B.T.M. is supported by the National Science Foundation Graduate Research Fellowship under Grant No. DGE-1144469. 

This research has made use of the Keck Observatory Archive (KOA), which is operated by the 
W. M. Keck Observatory and the NASA Exoplanet Science Institute (NExScI), 
under contract with the National Aeronautics and Space Administration.

These results made use of Lowell Observatory's Discovery Channel Telescope.
Lowell operates the DCT in partnership with Boston University, Northern Arizona University, the University
of Maryland, and the University of Toledo. Partial support of the DCT was provided by Discovery
Communications. 

This publication was made possible through the support of a grant from the John Templeton Foundation. 
The opinions expressed in this publication are those of the authors and do not necessarily reflect the 
views of the John Templeton Foundation.

The authors wish to recognize and acknowledge the very significant cultural role and reverence 
that the summit of Maunakea has always had within the indigenous Hawaiian community.  
We are most fortunate to have the opportunity to conduct observations from this mountain.

{\it Facilities:} \facility{DCT:DSSI}, \facility{Keck:I (HIRES)}, \facility{Keck:II (NIRC2)}

\end{document}